\documentclass[a4paper,twoside,reqno]{bjp}
\usepackage{color}
\usepackage{graphicx}
\usepackage{cite}
\usepackage{amssymb,amsmath,amscd,amsthm}
\usepackage{times}

\usepackage[bookmarks=false]{hyperref}
\hypersetup{%
    colorlinks=true,        
    linkcolor=blue,          
    citecolor=blue,         
    urlcolor=blue           
    }

\usepackage{geometry}
\allowdisplaybreaks

\begin{document}

\title{Two-time physics, Carroll symmetry and Jordan algebras}

\runningheads{A. Kamenshchik, A. Marrani, F. Muscolino}{Two-time physics, Carroll symmetry and Jordan algebras}

\begin{start}{%
\author{A. Kamenshchik}{1},
\author{A. Marrani}{2}
\author{F. Muscolino}{3}

\address{Dipartimento di Fisica e Astronomia ``A. Righi'', Universit\`a di Bologna and INFN, via Irnerio 46, 40126 Bologna, Italy }{1}
\address{School of Physics, Engineering and Computer Science, University of Hertfordshire, AL10 9AB Hatfield, UK}{2}
\address{Dipartimento di Matematica e Applicazioni, Universit\`a  di Milano Bicocca, via Roberto Cozzi 55, 20125 Milano, Italy}{3}

\received{Day Month Year (Insert date of submission)}
}

\begin{Abstract}
 We describe Carroll particles with nonzero energy (i.e., particles that remain at rest) within the
framework of two-time (2T) physics developed by Bars and collaborators. In a spacetime with one
additional time and one additional space dimension, one can gauge the phase-space symmetry that
exchanges generalized coordinates with their conjugate momenta, thereby unifying the description of
apparently different one-time systems. We develop both classical and quantum descriptions of the
Carroll particle arising from 2T physics, and explore links between the extended phase space of 2T
physics and Freudenthal triple systems constructed over a semisimple cubic Jordan algebra (the Lorentzian spin
factor).
\end{Abstract}

\begin{KEY}
Carroll symmetry,  Lorentzian spin factors, Freudenthal triple systems, two-time physics.
\end{KEY}
\end{start}


\section{Introduction}
Theories with extra spatial dimensions have become quite traditional since the times when they were put forward in the famous works by Kaluza and Klein.
Theories with more than one time dimension look much less intuitive and plausible. Nevertheless,
an impressive series of papers devoted to the so called two-time (2T) physics was produced by I. Bars and his co-authors  since 1996
(see the book \cite{book} and references therein).
Classical and quantum physics of simple systems such as non-relativistic particle, massive and massless relativistic
particles, harmonic oscillator, hydrogen-like atoms can be described in the framework of 2T-physics from a unifying point of view.
The language of 2T physics is also well adapted to field theories and to gravity.
However, relations between the 2T physics and the Carroll symmetry \cite{Carroll1,Carroll2,Carroll3}  had not been explored before.  This was done in our paper \cite{we}.

From the point of view of the 2T Physics,  usual physical systems living in a one-time world represent  projections from the spacetime with one additional temporal dimension one  additional spatial dimension.
These additional dimensions are  introduced to construct a new gauge theory, based on the localization   of the phase-space symmetry described by the symplectic group $Sp(2,\mathbb{R})$. Introduction of the gauge fields implies the presence of three first-class constraints $X\cdot X = 0, P\cdot P = 0, X\cdot P=0$, where $X$ and $P$ are
coordinates and momenta in the extended $2+d$-dimensional spacetime.
The usual physics in $1+(d-1)$-dimensional spacetime is obtained by means of a gauge-fixing procedure. Different gauge-fixings (or, in other words, different parametrizations
of the phase space coordinates $X$ and $P$ in terms of usual coordinates $x$ and momenta $p$) give different expressions for the   $1+(d-1)$-dimensional Hamiltonians and time parameters.

The Carroll Lie algebra and the Carroll group are obtained from the contraction of the Poincar\'e group by putting the velocity of light equal to zero \cite{Carroll1, Carroll2}.
The Carrol particle with non-zero energy should be always in rest.
The Carroll particle with zero  energy  is  always moving.
These two cases are  disconnected. This is because the Carroll boosts $K^i = x^i\frac{\partial}{\partial t}$ commute with the Hamiltonian
and do not change the energy, in contrast to the Lorentzian and Galilean boosts. In Section 2 we shall consider the classical Carroll particle in rest from the point of view of the 2T physics. In the third section we discuss its quantization. The fourth section is devoted to relations between 2T physics and Jordan algebras, extensively treated in \cite{we-future}. The last section contains concluding remarks.


\section{Carroll particle in two-time spacetime: classical theory}

The action for the Carroll particle can be represented as \cite{Carroll3}
\begin{align}\label{CarrollActionRest}
		S=-\int d\tau\left\{\dot{t}E-\dot{x}\cdot p - \lambda\left(E-E_0\right) \right\},
	\end{align}
	where $\tau$ is the proper time, $t$ is the physical time, $E$ represents the classical Hamiltonian and $x^i$ and $p^i$ are the space coordinates and the momenta, for $i=1,\dots, d-1$.
The quantity $E_0 \neq 0$  represents the rest energy of the Carroll particle and $\lambda$ plays the role of a Lagrange multiplier.
 We would like to obtain this action from the 2T action
 \begin{equation*}
 S = \int d\tau (P_A\dot{X}^A - A_{ij}Q^{ij},
 \end{equation*}
 where $A_{ij}$ are the gauge fields, playing the role of Lagrange multipliers while the constraints are
 $Q^{11} = X\cdot X, Q^{12} = X\cdot P, Q^{22} = P\cdot P$.
 Let us introduce the light-cone coordinates	
		\begin{align*}
			X^+=\frac{1}{2}\left(X^{1'}+X^{0'}\right),\quad X^-=\frac{1}{2}\left(X^{1'}-X^{0'}\right).
	\end{align*}
Choosing the two-time coordinates and momenta as in \cite{we}
\begin{eqnarray*}
&&X^+=E_0t,\\
&&X^-=\frac{x_ip^i}{E_0}+\frac{t}{E_0}\left(E-E_0+\frac{p_ip^i}{2}\right),\\
&&X^0=\sqrt{x_ix^i},\\
&&X^i = x^i+tp^i,\\
&&P^+=E_0,\\
&&P^-=\frac{1}{E_0}\left(E-E_0+\frac{p_ip^i}{2}\right),\\
&&P^0=0,\\
&&P^i=p^i,
\end{eqnarray*}
we obtain the action \eqref{CarrollActionRest}, provided $E=E_0$, which is obtained by solving the constraints $Q^{ij} = 0$.

\section{Carroll particle in two-time spacetime: quantum  theory}	

The commutation relation for the position and momentum operators in the standard $d-1$-dimensional space  are
	\begin{align*}
		[x^i,p^j]=i\ \delta^{ij}.
	\end{align*}
 Upon quantization, operator ordering becomes an issue.
All the operators should be Hermitian, but this requirement is not sufficient. We have to resort to the covariant quantization in  the 2T spacetime.
 The quantum generators $L^{MN}$ of $SO(2,d)$ must satisfy the Lie algebra under commutators. The generators of the Lorentz group $SO(2,d)$, $L^{MN}$ which become operators should constitute the Lie algebra with respect to the commutators.
This requirement also does not define the ordering in the quantum generators in a unique way and one should use also the properties of the Casimir operators of the unitary representations of both the groups  $SO(2,d)$ and\footnote{%
 We will henceforth use the physicists' notation of symplectic groups :
namely, $Sp(2,\mathbb{R})$ is the split real form of the Lie group whose
algebra is (in the usual Cartan's notation) $\mathfrak{c}_{1}$, when
considered over the complex numbers.}
 $Sp(2,{\mathbb R})$.
The constraints play the role of the generators of the symmetry with respect to the $Sp(2,{\mathbb R})$ group.
They should be applied to  the acceptable quantum states of the system  according to the prescription of the Dirac quantization of systems with first-class constraints
\begin{equation*}
Q|\Psi\rangle = 0.
\end{equation*}
The same should be valid also for the Casimir operators.
In paper \cite{Bars}
 this technique was implemented to reproduce the quantization scheme and the spectrum for the hydrogen-like atom.
If we manage to fix the ordering in the generators of $SO(2,d)$ group at $\tau=0$ , then the same ordering will be conserved.
Our parametrization of the variables $X^M,P^M$ at $\tau=0$ coincides with that used for the  description of the hydrogen atom in \cite{Bars}
provided we have already put $E=E_0$.
It is  amazing because these physical systems  are quite different and their actions are also different.

We can use this fact to quantize our Carroll particle at rest.
It does not mean that we shall obtain the discrete spectrum.
The combination of the squared momentum and the inverse radius is not connected with the Hamiltonian.
The momentum is not connected with the velocity  (which is equal to zero).
What is the role of the momentum?
It enters into the commutation relations  and, hence, the  Heisenberg inequality of uncertainties
	\begin{equation*}
	\Delta x^i\cdot \Delta p^j \geq \frac12\delta^{ij}
	\label{Heis}
	\end{equation*}
	is valid.
In contrast to the standard non-relativistic quantum mechanics,
	we can choose the quantum states with a dispersion of the coordinate $\Delta x$ as small as we wish,
	because the growth of the dispersion of the momentum $\Delta p$ is not important.
	Thus, a particle can be localized with an arbitrary high precision.

 \section{Two-time physics and Jordan algebras}

 The choice of the parametrizations of the coordinates and momenta in the 2T spacetime looks as some kind of the
 craftsman work. It would be interesting to find a  general algebraic structure behind the 2T spacetime.
 Perhaps, some rather abstract constructions connected with Jordan algebras can play this role (see \cite{we-future} for further details). Jordan algebras
 \cite{Jordan,Jordan1} were invented with an intention to create a new mathematical  apparatus for quantum theory.
 They did not  fulfill these expectations, but instead have found a lot of other applications in both mathematics and physics \cite{Jordan-book}.
 A  (real) Jordan algebra  $(\mathfrak{J},\circ)$  is a
vector space defined over a ground field $\mathbb{F}$ (in our case $\mathbb{R}$)
equipped with a
bilinear product $\circ$ which is commutative and  power-associative,  satisfying
\begin{equation*}
\begin{split}
X\circ Y& =Y\circ X, \\
X^{2}\circ (X\circ Y)& =X\circ (X^{2}\circ Y),\quad \forall \ X,Y\in
\mathfrak{J}.
\end{split}
\label{eq:Jid}
\end{equation*}
A cubic Jordan algebra is endowed with a cubic form $N:\mathfrak{J}\rightarrow \mathbb{R}$, such that
$N(\lambda
X)=\lambda ^{3}N(X),\quad \forall \ \lambda \in \mathbb{R},\ X\in \mathfrak{J%
}$. Besides, an element, named base point  $c\in \mathfrak{J}$ exists, satisfying $N(c)=1$.
 The definition of the cubic norm permits  to construct a cubic Jordan algebra, in which  all the
properties of the Jordan algebra are determined by the cubic form itself.
We are interested in a special class of the cubic Jordan algebras, called pseudo-Euclidean spin-factors.
 Having a $(m+n)$-dimensional pseudo-Euclidean spacetime $\Gamma_{m,n}$ we can construct a cubic Jordan algebra $
\mathfrak{J}=\mathbb{R}\oplus \Gamma_{m,n}$ with the cubic norm
\begin{equation*}
N_{3}(X=\xi \oplus \gamma ):=\xi \gamma ^{a}\gamma ^{b}\eta
_{ab}.
\end{equation*}
it is important that to transform a Minkowski spacetime $\Gamma_{1,d-1}$ into a cubic Jordan algebra, we should  add to this spacetime an additional spatial dimension.

Starting from a cubic Jordan algebra $\mathfrak{J}$, we can construct a Freudenthal
triple system \cite{Freudenthal}, which   is defined as the vector space
\begin{equation*}
\mathfrak{F}(\mathfrak{J}):=\mathbb{R}\oplus \mathbb{R}\oplus \mathfrak{%
J\oplus J}.
\end{equation*}
An element $\mathbf{x}\in \mathfrak{F}(\mathfrak{J})$ can  formally be
written as a \textquotedblleft $2\times 2$ matrix\textquotedblright\ :
\begin{equation*}
\mathbf{x}=%
\begin{pmatrix}
x & X \\
Y & y%
\end{pmatrix}%
,\text{\ }x,y\in \mathbb{R},~X,Y\in \mathfrak{J}.
\end{equation*}
A Freudenthal triple system is endowed with a non-degenerate  symplectic bilinear form, a quartic  invariant, and a trilinear triple product.
It is important  to note that, when constructing a Freudenthal triple system from the pseudo-Euclidean spin  factor, one algebraically doubles degrees of freedom from configuration to phase space.
\begin{equation*}
\underbrace
{\mathbb{R}\oplus (\mathbb{R}\oplus \Gamma_{1,d-1})}
_{coordinates}
\oplus \underbrace{\mathbb{R}\oplus (\mathbb{R}\oplus \Gamma_{1,d-1})}_{momenta}.
\end{equation*}
Besides, we can try to use the invariants obtained from the Freudenthal triple system structure  to classify different gauge-fixings in the two-time world.

\section{Conclusions}

We have found such a parametrization of the phase space variables in two-time spacetime, which permits to describe a Carroll particle in rest in the one-time spacetime.
In quantum theory we have seen an amusing correspondence between our parametrization and that used for the description and quantization of the hydrogen atom.
 The case of the always moving particle (Carroll tachyon) is more complicated and it is under study.
The most interesting line of research is probably an ``algebraic'' one: we hope to push forward the analysis of the relations between Jordan algebras and Freudenthal triple systems  and 2T physics to understand better how one can find different one-time worlds hidden inside of the two-time spacetimes.

\section*{Acknowledgements}
A.K. is grateful to the organizers of the conference QTS-13 in Yerevan, 2025 for the opportunity to give a talk.


\end{document}